\documentclass[conference]{IEEEtran}
\IEEEoverridecommandlockouts
\usepackage{cite}
\usepackage{amsmath,amssymb,amsfonts}
\usepackage{algorithmic}
\usepackage{graphicx}
\usepackage{textcomp}
\usepackage{xcolor}
\usepackage{xspace}
\usepackage[shortlabels]{enumitem}
\usepackage{url}
\usepackage{multirow}
\usepackage{booktabs} % For formal tables

\newcommand\urlBenchmark{https://gitlab.com/learnERC/DataLossRepository\xspace}
    \newcommand\NameFileHtml{Data\_Loss\_Apps.html\xspace}

\def\BibTeX{{\rm B\kern-.05em{\sc i\kern-.025em b}\kern-.08em
    T\kern-.1667em\lower.7ex\hbox{E}\kern-.125emX}}

\begin{document}

\title{A Benchmark of Data Loss Bugs for Android Apps}
%\thanks{This work has been partially supported by the EU H2020 Learn project, which has been funded under the ERC Consolidator Grant 2014 program (ERC Grant Agreement n. 646867) and the GAUSS national research project, which has been funded by the MIUR under the PRIN 2015 program (Contract 2015KWREMX).}
\author{\IEEEauthorblockN{Oliviero Riganelli, Marco Mobilio, Daniela Micucci, Leonardo Mariani}
\IEEEauthorblockA{\textit{Department of Informatics, Systems and Communication} \\
\textit{University of Milano - Bicocca}, Milan, Italy\\
\{oliviero.riganelli, marco.mobilio, daniela.micucci, leonardo.mariani\}@unimib.it}
}

\maketitle

\begin{abstract}
Android apps must be able to deal with both \emph{stop event}s, which require immediately stopping the execution of the app without losing state information, and \emph{start event}s, which require resuming the execution of the app at the same point it was stopped. Support to these kinds of events must be explicitly implemented by developers who unfortunately often fail to implement the proper logic for saving and restoring the state of an app. As a consequence apps can lose data when moved to background and then back to foreground (e.g., to answer a call) or when the screen is simply rotated. These faults can be the cause of annoying usability issues and unexpected crashes.

This paper presents a public benchmark of 110 data loss faults in Android apps that we systematically collected to facilitate research and experimentation with these problems. The benchmark is available on GitLab and includes the faulty apps, the fixed apps (when available), the test cases to automatically reproduce the problems, and additional information that may help researchers in their tasks.
\end{abstract}


% !TEX root =  Main.tex
\section{Introduction} \label{sec:introduction}

Interactions with Android apps may produce many \emph{stop event}s, which require suspending the execution of the apps, and \emph{start event}s, which require resuming the execution of the apps. Many frequent situations may trigger these kinds of events: switching between applications, rotating the screen, switching between windows in the same app, and so on. 

%For instance, switching from an app to another one produces a stop event for the first app and a start event for the second one. Answering a phone call causes the suspension of the execution for the current app, which has to handle a stop event. When the call ends, the execution of the app can be resumed and a start event is handled. Finally, simply rotating the screen of a device requires first suspending the execution of the app and then resuming its execution, requiring the app to handle a stop event followed by a start event in addition to handling the new orientation of the screen.

Handling stop and start events requires implementing the logic necessary to save the state of the app, when the stop event is received, and to restore the state of the app, when the start event is received~\cite{Android_Lifecycle}. This logic must be implemented within the callbacks methods that are invoked by the Android framework when stop and start events occur. For instance, when the user rotates the screen, the Android framework first invokes the \texttt{onSaveInstanceState()} callback method to save the state of the current activity, then destroys the activity, adjusts the orientation of the screen, reloads the activity, and finally recovers the saved state by invoking the \texttt{onRestoreInstanceState()} callback method.
There are indeed several possible causes for data loss problems. For example, the callback methods involved in this process might be missing or might be implemented incorrectly; or framework upgrades may change the generation of the callbacks breaking the logic of the app~\cite{Riganelli:DataLoss:2016}. 
%
%This logic must be executed from the callback methods of the Android components that constitute the application. 
Saving and restoring state information might be tricky because each app requires a specific implementation that depends on the nature of the data that must be saved and resumed. In fact, developers can easily miss to properly save or resume some fields for some components and introduce data loss faults in their apps~\cite{Hu_EED_2014,Adamsen_SEA_2015,Amalfitano:Rotation:STVR:17}.

%On the first place a 
A data loss fault causes some of the program variables to lose their values, which are replaced by default values once the app is resumed (e.g., numeric data types are assigned with 0 and objects are assigned with \texttt{null}). This might have a range of consequences depending on the variables that are assigned with incorrect values. In the best case, users may just loose the values they entered into the app, %(e.g., the app restarts with all empty fields or from the home activity once it is resumed). 
in other cases, variables with wrong values might be the cause of incorrect computations and crashes.  

%Developing techniques that are able to detect, locate, and repair these faults requires the availability of an extensive set of apps and faults covering a variety of cases. 

In addition to generic approaches for the detection, localization, and repair of Android faults~\cite{Machado:MZoltar:IWSD:2013,Gao:SymbolicTesting:ASE:2018,Seams17,RV17,Machiry:Dynodroid:2013,Gazzola:Repair:TSE:2017}, it is important to design techniques that can help developers with data loss problems~\cite{Hu_EED_2014,Adamsen_SEA_2015,Zaeem:DataLossOracle:2014,Riganelli:DataLoss:2016}. %For instance, these tools can autonomously repair the application or can support the developer by highlighting the problem.
To be able to experiment with a wide class of faults related to data loss and compare techniques, it is important to exploit publicly available data sets that include extensive sets of apps and faults, covering a variety of cases.

%In the last few years, several techniques have been proposed to detect, locate, and rapir Android bugs~\cite{Hu_EED_2014,Adamsen_SEA_2015,Riganelli:DataLoss:2016,Seams17,RV17,Machiry:Dynodroid:2013,Zaeem:DataLossOracle:2014,Gazzola:Repair:TSE:2017}. Unfortunately, techniques are often experimented on different sets of apps and bugs, making comparison extremely hard. 

%Developing such techniques requires the availability of an extensive set of apps and faults covering a variety of cases. 

This paper presents the result of our effort in the creation of a public repository of 110 real data loss faults affecting 48 Android apps. The repository is designed to facilitate the reproduction of the data loss faults: it includes the faulty apps, the fixed apps (when available), and test cases that fail on the faulty apps but pass on the fixed apps that any third party can use to automatically reproduce the data loss problems. The repository includes additional information that may help researchers with their tasks, such as the version of both the emulator and the Android API that we used to reproduce the faults. The interested researchers and professionals can access the repository at the following url: \urlBenchmark and can use our benchmark to evaluate their techniques against data loss defects. We expect our benchmark to support and facilitate the definition of methods and techniques to detect and fix data loss faults in Android apps.

%In the last few years, several static and/or dynamic analysis techniques have been proposed to detect, locate and rapir android bugs~\cite{Hu_EED_2014,Adamsen_SEA_2015,Riganelli:DataLoss:2016,Seams17,RV17,Machiry:Dynodroid:2013,Zaeem:DataLossOracle:2014}. Unfortunately, techniques are often experimented on different sets of apps and bugs, making comparison extremely hard. Interested researchers and professionals can use our benchmark to evaluate their techniques against data loss defects. We expect our benchmark to support and facilitate advancements in detecting and fixing data loss faults in Android apps.

The rest of the paper presents the methodology that we adopted to create the benchmark (Section~\ref{sec:methodology}), describes the artefacts that are part of the benchmark (Section~\ref{sec:benchmark}), illustrates how the benchmark can be used (Section~\ref{sec:usage}), describes challenges, limitations, and improvements (Section~\ref{sec:challenge}), and provides final remarks (Section~\ref{sec:conclusion}).
% !TEX root =  Main.tex
\section{Methodology} \label{sec:methodology}
The methodology that we used to build our benchmark of apps affected by data loss faults consists of four main steps:
\begin{enumerate}
	\item \textit{Selection of the eligible apps}. In this step, we identify the repositories and data sets of apps affected by data loss faults that we consider to create our benchmark.
	\item \textit{Identification of the apps that satisfy the reproducibility requirements}. In this step, we filter out the apps that do not satisfy our reproducibility requirements from the overall set of apps selected in the previous step.
	\item \textit{Compilation and execution of the apps}. In this step, we work on the selected apps to make sure they can be compiled and executed, which are necessary conditions to reproduce failures.
	\item \textit{Reproduction of the data loss faults}. In this last step, for each fault we implement an automatic test case that interacts with the graphical user interface of the application to reproduce it. 
\end{enumerate}

We describe each step in detail below.

\subsection{Selection of the eligible apps}
To create the benchmark, we considered the Android apps with one or more data loss faults available on F-Droid ({\small \url{https://f-droid.org/en/}}) in early June 2018. F-Droid is a well-known software repository that at the time of our search contained 1,420 free and open-source apps for the Android platform. In order to select significant and mature app projects, we applied the following selection criteria:
\begin{itemize}
\item \emph{Traceable}: the app should have a public version control and issue tracking system.
\item \emph{Popular}: The app should have more than 10,000 downloads on Google Play ({\small \url{https://play.google.com/}}). 
\item \emph{Maintained}: The app should contain more than 100 code revisions.
\item \emph{Non-Trivial}: The app should contain at least 1,000 lines of Java code.  
\end{itemize}

The application of these criteria generated a list of 428 apps from F-Droid whose software repositories are mainly hosted on GitHub ({\small \url{https://github.com/}}). We thus analyzed the project history of these apps to discover the potential data loss faults by performing keyword search in commits and bug reports. In order to reduce the risk of losing real data loss problems, we leverage on general keywords that are  \emph{data}, \emph{loss}, \emph{landscape}, \emph{save}, \emph{rotate}, \emph{screen}, \emph{portrait}, and \emph{restore},  with the aim of maximizing the coverage of data loss issues. We also used \emph{onSaveInstanceState} and \emph{onRestoreInstanceState} as keywords since they are the names of the callback methods implemented to save and restore the app data.   For each general keyword, we used the appropriate word forms, such as conjugations and declensions. For example, the keyword \emph{rotate} was also searched for \emph{rotation}, \emph{rotated}, and \emph{rotating}. The search resulted in more than 2,500 cases of potential data loss faults.

%
%\begin{table}
%\centering
%\caption{Keywords for Mining Commits and Bug Report}
%\label{tab:keywords}
%\begin{tabular}{|c|c|c|c|c|c|} 
%\hline
%\multicolumn{2}{|l|}{onSaveInstanceState} & \multicolumn{2}{l|}{onRestoreInstanceState} & Data     & Loss    \\ 
%\hline
%Landscape & Save                          & Rotate & Screen                             & Portrait & Restore  \\
%\hline
%\end{tabular}
%\end{table}

\subsection{Identification of the apps that satisfy the reproducibility requirements}
In this step, we manually analyze the commits and the bug reports selected in the previous step to make sure that the occurrence of the keywords is not incidental and that the data loss faults are actually present. Since the search operation was conservatively inclusive, for instance selecting every bug report that simply mentions the word \emph{rotate}, we manage to quickly discard most of the entries by inspecting the bug reports, and only a few cases required extensive checks. Our manual analysis finally confirmed the presence of 168 bugs affecting 82 different apps as actual cases of data loss faults.

%Nearly one out of five apps (19.2\%) were affected by data loss, which shows the pervasiveness of data loss issues in mobile apps. 
Out of 428 apps that satisfy our requirements about maturity and significance, 82 apps (19.2\%) were affected by at least a data loss fault, which shows the pervasiveness of data loss issues in mobile apps. 
Data loss faults may be not trivial to localize and repair as witnessed by discussions and commits: faults require on average an online discussion of 3 comments to be clarified (15 comments in the worst case and 1 comment in the best case); fixes can spread over multiple methods and files, in fact they required the modification of at least two files in 68\% of the cases, and they required the modification of 44 lines of code on average (1 line of code in the best case, and 1,856 lines of code in the worst case). The studied faults required from 1 day to almost a year to be fixed.
% 153 bug reports and 74 fix commits, for a total of 168 unique cases of potential data loss faults affecting 82 different apps. are actual cases of data loss

To facilitate the compilation and reproduction tasks, we considered only faulty apps developed in Android Studio ({\small \url{https://developer.android.com/studio/}}). This produces a negligible reduction of our data set, in fact of these 168 cases, only 16 apps affected by 22 data loss faults were not developed in Android Studio.  We thus ended up with 146 data loss faults affecting 78 releases of 66 apps.

\subsection{Compilation and execution of the apps}
In order to reproduce data loss faults, it is mandatory that each app release can be \emph{executed} and that the executed app is consistent with the available source code. We thus worked on the compilation and execution of each app release performing the following steps.

\begin{enumerate}
\item \label{StepDownload} We downloaded the source code of the 78 app releases affected by the data loss faults from the public version control system of the app linked by F-Droid.  When available, we also downloaded the app releases containing the fixes.

\item \label{StepImport} We imported all the downloaded projects into the Android Studio IDE. 

\item \label{StepCompile} We compiled the code of the apps.
\begin{itemize}
\item We first checked the presence of compilation instructions that can help us with this task.
\item We then compiled each app using the Android Studio IDE. 
\end{itemize}

\item \label{StepFixErrors} When experiencing compilation errors, we tried to fix the app. In total we experienced compilation problems for 39 app releases. We managed to successfully fix the compilation problems for 16 of them by applying one or more of these actions.
\begin{itemize}
\item Change the version of the Gradle ({\small \url{https://gradle.org}}) plugin in the dependencies (in most cases with version 2.3.3).
\item Change the Gradle version to the Gradle Wrapper if the Android Gradle plugin and Gradle versions are not compatible (in most cases with version 3.3).
\item Update the SDK Build Tools revision in case the version is not compatible.
\item Import the Google Maven repository ({\small \url{https://mvnrepository.com/artifact/com.google}}).
\item Download the configuration file google-services.json from Firebase ({\small \url{https://firebase.google.com/}}) and add to the app directory.
\item Update some configurations if obsolete (e.g., 'compile' replaced with 'implementation').
\item Configure a build.gradle file in case it is not already configured.
\end{itemize}

\end{enumerate}

%After step (\ref{StepDownload}) we managed to successfully download the source code of 78 buggy releases of the 66 selected apps. Step (\ref{StepImport}) did not suppress any app because all the 78 releases have been successfully imported into their respective IDE. After step (\ref{StepCompile}), 39 app releases passed the compilation stage without requiring any fix, while 39 failed to compile. We fixed 16 of these apps (\ref{StepFixErrors}) while the remaining 23 buggy releases have been dropped.
%The fixes that we implemented are distributed as follows (note that we implemented multiple fixes for some apps): XXXTotalAppsKOCompilationFixedVersion app releases had an incompatibility between the Android version declared in the configuration file and the one in the project properties; XXXTotalAppsKOCompilationFixedExternalLib app releases required external libraries; XXXTotalAppsKOCompilationFixedXML presented errors in the XML files; and XXXTotalAppsKOCompilationFixedJava presented Java errors. 

The main problem with unfixed app releases is outdated dependencies that were difficult or impossible to satisfy. This activity finally resulted in 55 app releases of 49 apps that have been successfully compiled. 

We then executed all the 55 app releases that we compiled. In particular, we ran 53 app releases on an emulated Google Nexus 5 with Android 6.0, and  we ran 2 app releases on an emulated Google Nexus 5 with Android 5.1. This resulted in a total of 116 data loss faults affecting 55 releases of 49 apps that can be potentially reproduced by using the apps.

\subsection{Reproduction of the data loss faults}
This is the last and most difficult step of our process, that is, the \emph{reproduction} of the data loss faults and the implementation of the \emph{automatic test case}s that reveal the faults.
To achieve both these objectives, we performed the following steps.

\begin{enumerate}
\item We extracted \emph{information useful to reproduce the data loss fault}. In particular, we analyze the app to identify the variables whose values are lost and the conditions that cause the data loss. When possible, we started our investigation from the bug report. If the bug report was not available, we started from the comments in the commits and the code that fixes the discovered bugs to infer what values are lost and under what conditions. Overall, we successfully analyzed 116 data loss faults.

\item We worked on the \emph{identification of the sequence of end-user operations} (i.e., interactions with the user interface) that made the app fail because of the data loss. 
In 6 cases, it was impossible to understand if the data loss indicated in the commit operation could be feasibly reproduced. We however managed to reproduce 110 out of the 116 data loss faults that have been selected. 

\item We implemented an \emph{automatic test case} for each data loss we manually reproduced.  We used Appium ({\small \url{http://appium.io}}) and the Genymotion emulator ({\small \url{https://www.genymotion.com}}) to implement the test case.

\item In all the cases the reproduction of the data loss implied immediately visibile effects on the faulty app. We managed to implement an automatic \emph{oracle} in the vast majority of the cases (98 out of 110). The oracle checks the behavior of the app and makes the test fail when the data loss is reproduced (e.g., we verify that after a rotation the text written in a form is not lost). The few cases without an automatic oracle depend on Appium not being able to read the properties of the widget presenting the data loss problem.

%implemented a test case that reproduces the data loss problem for 98 of the cases. For 12 cases, we failed to implement an oracle because it was impossible to extract  
%test cases include an oracle, while 12 test cases have no oracle. In the latter cases the loss of data is visible to the user, but we have not managed to find a way to specify the oracle for a specific view that loses the data in these cases.
\end{enumerate}

% !TEX root =  Main.tex
\section{Data Loss Benchmark} \label{sec:benchmark}

%how data is organized (folders, content, etc)
%originality: claim no other benchmark like this is available
%add benchmark stats

The benchmark consists of 110 reproducible data loss faults present in 54 releases of 48 different apps and is hosted on GitLab at the following address:\linebreak {\small \urlBenchmark}. 

The root of the project includes a folder for each Android app affected by at least a data loss. The name of the folder is the same as the name of the app. The folder of an app further includes folders with the faulty releases and the fixed releases.
The root of the project also hosts the \NameFileHtml file that reports information about every data loss fault that has been reproduced. In particular, each data loss fault is associated with the following information (see Table~\ref{tab:benchmark} for two sample entries):

%
%The following entry describe each the resource leak present th the benchmark.
\begin{itemize}
\item General Info
	\begin{itemize}
	\item \emph{App Name}: the name of the app affected by the data loss.
        	\item \emph{Category}: the category to which the app belongs to.
	\item \emph{Issue Report}: the identifier of the bug report that describes the data loss.
        \end{itemize}
        
\item Faulty app info     
	\begin{itemize}
	%		\item \emph{Buggy Version}: the identifier of the buggy version in the version control system of the app.
	\item \emph{Faulty Version}: the release of the app affected by the data loss fault in the version control system of the app.
	\item \emph{Faulty Apk}: the name of the apk file corresponding to the \emph{Faulty Source Code} that is available in our benchmark in the folder of the app.
	\item \emph{Faulty Source Code}: the name of the zip file containing the source code of the faulty version that is available in our benchmark in the folder of the app. It includes the manual fixes that we implemented to compile and execute the app.
	\item \emph{Faulty Activity}: the name of the Activity class that is affected by the data loss issue.
			\end{itemize}

\item Fixed app info      
	\begin{itemize}
	\item \emph{Fixed Version}: the identifier of the app version with the fix as it appears in the version control system of the app.
	\item \emph{Fixed Apk}: the name of the apk file corresponding to the \emph{Fixed Source Code} that is available in our benchmark in the folder of the app.
		\item \emph{Fixed Source Code}: the name of the zip file containing the source code of the fixed version that is available in our benchmark in the folder of the app. It includes the manual fixes that we implemented to compile and execute the app.
			\end{itemize}
	
\item Execution info
	\begin{itemize}	
	\item \emph{Test Case}: the name of the zip file that contains an Appium automatic test case that reveals the bug. It is available in our benchmark in the folder of the app.
	\item \emph{Oracle}: it indicates if the test case includes an oracle.
		\item \emph{Tested API}: it indicates the API version of the emulator we used to reproduce the failure.
		\item \emph{Target API (Comp - Min)}: the highest API level against which the application was designed, the version of the Android API with which the app is compiled, and the minimum API level with which the app is compatible, respectively.

	\end{itemize}
\end{itemize}

\begin{table}[h]
%\resizebox{\textwidth}{!}{%

  \centering

\scriptsize
\caption{Two samples of data losses in the benchmark.}
\label{tab:benchmark}

\begin{tabular*}{\columnwidth}{@{\extracolsep{\stretch{1}}}*{3}{l}@{}}%[t]{llll}
%\toprule
\emph{\textbf{General info}}\\
\cmidrule{1-1} 
\textbf{App Name} & \textbf{Category} & \textbf{Issue Report}\\
\midrule
Amaze File Manager & Tools & Issue 1034\\ 
OpenTasks & Productivity & Issue 658\\
\midrule
\end{tabular*}\\ \qquad

%%%%%%%%%%%%%%%%%%

\begin{tabular*}{\columnwidth}{@{\extracolsep{\stretch{1}}}*{5}{l}@{}}%[t]{llll}
 \\
\emph{\textbf{Faulty app info}} \\
\cmidrule{1-1} 

\textbf{Faulty Version} & \textbf{Faulty Apk} & \textbf{Faulty Source Code} & \textbf{Faulty Activity}\\
\midrule
v3.1.0-beta.1 & v3.1.0-beta.1.apk & v3.1.0-beta.1.zip & MainActivity \\
v1.1.13 & v1.1.13.apk & v1.1.13.zip  & EditTaskActivity\\
\midrule
\end{tabular*}\\

%%%%%%%%%%%%%

\begin{tabular*}{\columnwidth}{@{\extracolsep{\stretch{1}}}*{3}{l}@{}}%[t]{llll}
 \\
\emph{\textbf{Fixed app info}} \\
\cmidrule{1-1} 
\textbf{Fixed Version} & \textbf{Fixed Apk} & \textbf{Fixed Source Code}\\
\midrule
884c16c & 884c16c.apk & 884c16c.zip \\
N/A & N/A & N/A \\
\midrule
\end{tabular*}\\

\begin{tabular*}{\columnwidth}{@{\extracolsep{\stretch{1}}}*{4}{l}@{}}%[t]{llll}
 \\
\emph{\textbf{Execution info}} \\
\cmidrule{1-1} 
\textbf{Test Case} & \textbf{Oracle} & \textbf{Tested API} & \textbf{Target API (Comp-Min)}   \\
\midrule
TestCase 1034 & yes  & 22 & 25 (25 - 14) \\
TestCase 658 & yes  & 23  &  25 (25 - 15) \\
\midrule

\end{tabular*}\\ 
%}

\end{table}

% !TEX root =  Main.tex
\section{Benchmark Usage} \label{sec:usage}

%ideas for future research questions that could be answered using the data set,
%ideas for further improvements that could be made to the data set, and
%any limitations and/or challenges in creating or using the data set.
The benchmark can be used to study the effectiveness of techniques for identifying, analyzing, and fixing faults in Android apps. 
The first step normally is reproducing the available data loss faults locally. To do this, the recommended procedure is the following one:
\begin{enumerate}
\item Clone the benchmark repository in the target computer.
\item If not already present, install Appium (we used version 1.3.1) and its dependencies (the JDK and Android SDK).
\item Configure and run Appium server.  
The test cases assume that the app under test runs on the same machine of the Appium client and that the Appium server is available on the port 4723. If it is not the case, the test case must be changed accordingly. 
\item Set \textit{ANDROID\_HOME} and \textit{adb} in the environment variables.
\item Start the emulator/device according to the attribute \textit{MobileCapabilityType.}\textit{VERSION}
    specified in the test case code.
\item Install the apk of the app, launch Appium, and finally launch the test case to reproduce the data loss.
\end{enumerate}

The dataset contains all the precompiled apks of the available apps, which were built starting from the releases specified in the Data\_loss\_Apps.html file of the dataset.

\smallskip

The benchmark can be potentially used to answer a number of research questions. Indeed it allows to study the \emph{effectiveness} and \emph{efficiency} of a range of analysis and testing techniques for Android apps. In particular, the benchmark can be used to assess \emph{static analysis} techniques, since it makes the source code of the faulty and fixed apps available; to assess \emph{dynamic analysis} techniques, since it makes the automatic test cases for failure reproduction available; and \emph{testing} techniques, since it makes the faulty apps available. Moreover, it can be exploited to study the \emph{evolution} of these bugs, since the faulty and fixed versions of the apps are explicitly mapped to the corresponding revisions in GitHub.

\section{Challenges, Limitations and Improvements} \label{sec:challenge}

%Although there exist other benchmarks of faults in Android apps~\cite{Mitra:Vulnerability:Promise:2017,doi:10.1002/spe.2672}, our benchmark originally includes data loss faults equipped with automatic test cases and both source and compiled apps with faults and fixes.

Although there exist benchmarks of vulnerability faults~\cite{Mitra:Vulnerability:Promise:2017} and resource leaks~\cite{doi:10.1002/spe.2672} in Android apps, our benchmark \emph{originally} includes data loss faults equipped with automatic test cases and both source and compiled apps with faults and fixes.

To create this benchmark we faced several \emph{challenges}. For instance, we performed a significant magnitude of manual work by manually analyzing thousand of reports and commits, %We cannot exclude that a small number of entries have been misclassified as not being related to data loss faults. We have however confidence on our analysis, and the high rate of apps affected by data loss faults that we reported confirms that the analysis has been carefully performed. 
we fixed several tricky compilation problems and packaged the apps in artefacts ready to be used, and
we invested significant effort in carefully reproducing all the faults, de facto confirming them, and implementing automatic test cases that can be inexpensively executed by any third party.

A \emph{limitation} of our benchmark is that projects have been compiled with the Android Studio IDE, and users who want to use a different IDE may have to fix some compilation problems. %, as we did for some of the apps. 
This is anyway a marginal problem since it is not an obstacle to the usage of the artefacts (e.g., the apk and the test cases) that are part of the benchmark.

The benchmark is already quite broad including 110 data loss faults. It could be however further improved including other faults and apps extracted from other official repositories. Another possible improvement is the creation of artefacts that can be executed with virtually no configuration effort, such as making containers with the Android emulators already set up available. However, combining multiple virtualization levels can be problematic, and our benchmark requires only a few operations to be used.     
%:
% !TEX root =  Main.tex
\section{Conclusion} \label{sec:conclusion}
Android apps are frequently affected by data loss faults (19.2\% of the analyzed projects in our investigation). %that is, apps do not properly manage stop-start events causing annoying usability issues and unexpected crashes. 
It is therefore important to have techniques that can detect, localize, and repair these problems. To be able to experiment on a wide class of  data loss faults and assess testing and analysis techniques, we produced a publicly available data set that includes an extensive set of apps and faults. Interestingly our benchmark includes not only the faults, but also the apk and the automatic test cases to reproduce these bugs, delivering a total of 110 data loss faults. We expect our benchmark to facilitate progresses in detecting and repairing data loss faults in Android apps.

\smallskip

%19.2\% of the apps analyzed were affected by data loss, this demonstrates how this problem is able to escape to current testing practices~\cite{Rubinov:TestingAndroid:2018}. We therefore expect our benchmark to support and facilitate progress in detecting and repairing data loss in Android apps.

\begin{small}
\subsubsection*{Acknowledgements}
%\medskip
%\emph{Acknowledgements}
This work has been partially supported by the EU H2020 Learn project, funded under the ERC Consolidator Grant 2014 program (Grant Agreement n. 646867) and the GAUSS national research project, funded by the MIUR under the PRIN 2015 program (Contract 2015KWREMX).
\end{small}

\bibliographystyle{IEEEtran}
\bibliography{bibliography}

\end{document}